\documentclass[journal=apchd5,manuscript=article]{achemso}

\usepackage{chemformula} 
\usepackage[T1]{fontenc} 
\usepackage{amsmath}
\usepackage{graphicx}
\usepackage{multirow}

\author{Yihang Dan}
\affiliation{State Key Laboratory of Information Photonics and Optical Communications,
Beijing University of Posts and Telecommunications, Beijing}
\author{Tian Zhang}
\affiliation{State Key Laboratory of Information Photonics and Optical Communications,
Beijing University of Posts and Telecommunications, Beijing}
\email{ztian@bupt.edu.cn}
\author{Jian Dai}
\affiliation{State Key Laboratory of Information Photonics and Optical Communications,
Beijing University of Posts and Telecommunications, Beijing}
\author{Kun Xu}
\affiliation{State Key Laboratory of Information Photonics and Optical Communications,
Beijing University of Posts and Telecommunications, Beijing}

\title[An \textsf{achemso} demo]
  {Programmable Multifunctional Plasmonic Waveguide System based on 
  Coding Metamaterials and Inverse Design}

\abbreviations{IR,NMR,UV}

\begin{document}

\begin{tocentry}
\includegraphics{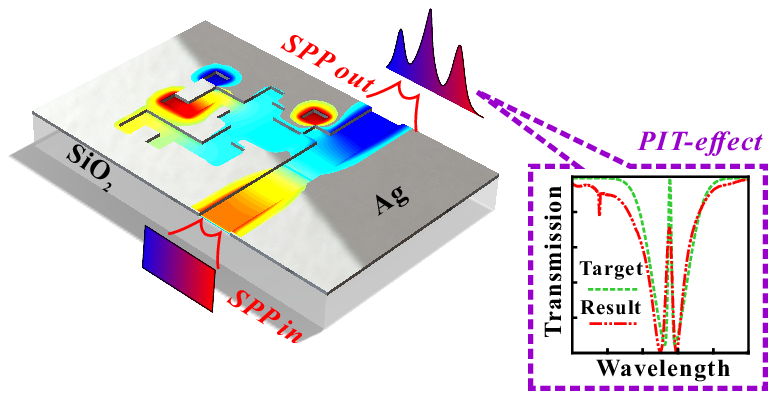}

Table of Contents Entry for \emph{Programmable Multifunctional Plasmonic 
Waveguide System based on Coding Metamaterials and Inverse Design} by
Yihang Dan, Tian Zhang, Jian Dai, and Kun Xu.

Plasmon-induced transparency-like effect in a metal-dielectric-metal waveguide system 
is realized based on metal coding metamaterials and inverse design. The metal coding 
metamaterials marked with green dotted line is optimized by gradient free methods.

\end{tocentry}

\begin{abstract}
  In this article, we propose a programmable plasmonic waveguide 
  system (PPWS) to achieve several different functions based on 
  metal coding metamaterials (MCMs) and inverse design technology. 
  There is no need to spend much time on considering the relation between 
  the function and the structure because the MCMs in the PPWS 
  are reprogrammable. In order to demonstrate the effectiveness of 
  the PPWS, we utilize it to achieve several filtering functions, 
  including bandstop and bandpass filters. The simulation 
  results exhibit that the performance of filters is improved based 
  on genetic algorithm, particle swarm optimization, multi-traversal 
  direct-binary search and simulated annealing. Especially, the bandwidth 
  and quality factor for the narrow-band filter can reach 6.5 nm and 200.5. 
  In addition to the simple filtering functions, some relatively complex 
  transmission characteristics can be obtained by using the PPWS, 
  such as plasmon-induced transparency-like effects. In conclusion, 
  genetic algorithm is considered as the most efficient inverse design 
  method for our system due to its less optimization time and 
  stable performance. In comparison with the previous works, 
  our proposed PPWS not only provides a general framework for 
  obtaining an effective, flexible and compact plasmonic device 
  but also shows the applications of inverse design on photonics devices.
\end{abstract}

\section{Introduction}

  Surface plasmon polaritons (SPPs) are especial electromagnetic waves 
  which occur and propagate at the interface of metal and dielectric 
  and they carry energy and information overcoming the diffraction limit
  \cite{Cunningham1974}. The appearance of SPPs provides a new way to manipulate light at 
  the nanoscale so it is considered to be the most potential way to 
  realize highly integrated optical system in the future \cite{ZAYATS2005}. 
  Since SPPs are discovered, researchers have studied different ways 
  to excite SPPs, such as grating excitation \cite{Hopper2002}, prism coupling 
  \cite{Otto1968,Kretschmann1968}, waveguide mode coupling \cite{Lavers1994}, 
  end-fire coupling \cite{Stegeman1983} and local excitation \cite{Hecht1996}. 
  The transmission of SPPs usually depends on special waveguide systems, 
  for example metal-dielectric-metal (MDM) waveguide \cite{Kocabas2009}, 
  insulator-metal-insulator waveguide \cite{Du2018}, hybrid waveguide \cite{Hua2017} 
  and so on. Due to their compact footprint and easy integration, 
  MDM waveguide systems have been widely applied in plasmonic filters \cite{Lin2008,Lu2010},  
  splitters \cite{Kwon2015,Yang2019}, optical switches \cite{JS2013},  
  modulators \cite{JS2013,Emboras2015,Hoessbacher2017}, absorbers \cite{Zhang2011}, 
  sensors \cite{Lu2012}, couplers \cite{Ruoxi2010,Virendra2019}, logic gates \cite{Liu2019} 
  and so on. However, in order to realize different functions, researchers often 
  spend much time on searching for and designing the nano-resonators 
  with different shapes \cite{Zhang2014,Liu2017surface,lai2018plasmonic}. 
  For example, the plasmon-induced transparency (PIT) effect 
  \cite{han2011plasmon,wang2012dispersionless}, which has applications 
  in refractive index sensors and slow light, is a hot spot in the field of SPPs. 
  In order to achieve the PIT effects in the transmission spectrum, various structures of waveguide systems 
  with different resonators have been proposed, such as ring resonators \cite{zhan2014slow}, 
  stub resonators \cite{chen2012multiple}, T-shaped resonators \cite{lu2011induced}, 
  H-shaped resonators \cite{chen2018tunable}, rectangular resonators \cite{han2015ultrafast}, 
  side-coupled multiple resonators  \cite{guo2014plasmon,liu2015tunable,lu2012plasmonic,he2014combined} and so on. 
  Obviously, it is cumbersome to construct different resonators for a special purpose. 
  It is expected to improve the design efficiency of photonics devices if 
  there exists a general framework, which can realize multiple functions and avoid 
  repetitive design works. It has been demonstrated that the coding metamaterials, 
  which consist of two types of unit cells, can provide a flexible and controllable platform by using a 
  field-programmable gate array \cite{cui2014coding}. The coding metamaterials are effectively 
  designed to achieve high-performance and easily integrated photonics devices 
  based on the inverse design technology \cite{Jia2018,shen2015broadband,xie2020broadband,Wang2020}. 
  However, it should be noted that the above waveguide systems generally 
  include dielectric coding metamaterials. There are few plasmonic waveguide systems 
  having programmable metal coding metamaterials (MCMs) to obtain multi-functions in a general framework.
  
  The inverse design technology based on optimization algorithms, 
  has been used to search for the appropriate distributions and structure 
  parameter of the MCMs. In general, there are three main ways to implement 
  the inverse design of photonics devices: gradient based methods, gradient 
  free methods and model-based methods \cite{molesky2018inverse}. 
  As a representative method of thegradient based methods, adjoint variable method (AVM) can optimize for the 
  linear and nonlinear optical devices, but it needs physical background to 
  derive the gradient of objective function \cite{hughes2018adjoint}. Model based methods, 
  such as artificial neural networks (ANNs) and random forests, 
  are also used for the inverse design of photonics devices \cite{zhang2019efficient}. 
  However, a lot of original data and labels are required to train 
  the model which constructs the physical relationship between 
  the structure parameters and corresponding optical responses. 
  The optimization performance is highly dependent on 
  the effect of pre-training model \cite{liu2018training}. In comparison to 
  the model-based methods, the gradient free methods, including search strategy and evolution strategy, 
  are simple and effective as they directly search for the optimal solution 
  by iteration and evolution strategies. Genetic algorithm (GA) and particle swarm 
  optimization (PSO) are two representative evolution algorithms, 
  which are inspired by the genetic inheritance and group cooperation \cite{Adaptation1992,kennedy2002particle}. 
  Although GA and PSO have been widely applied in the optimization and design of 
  photonic devices \cite{skaar1998a,kern2003a,chen2008synthesis,zhang2020efficient,shokoohsaremi2007particle,
  forestiere2010particle,lu2018particle,Kim2004}, they are easy to converge too early and fall into 
  local optimal solution in sometimes. Moreover, simulated annealing (SA) and
    direct-binary search (DBS) are traditional search algorithms which require 
  less time to converge \cite{Kim2004,Shen2015}. However, these two algorithms also easily 
  fall into local optimal solution because they are sensitive to the initial 
  states and optimization parameter settings. For specific inverse design tasks 
  with different frameworks, the performance of these algorithms are very different. 
  It is necessary to select the most suitable algorithm according to the actual performance. 
  
  In this paper, we propose a programmable plasmonic waveguide system (PPWS) 
  based on MCMs and the inverse design to obtain several different functions. 
  Four optimization algorithms, including GA, PSO, SA and DBS, are used to 
  design for the MCMs in the PPWS to realize multiple optical filtering functions, 
  such as narrowband band stop filter (NBSF), broadband band stop filter (BBSF), 
  narrowband band pass filter (NBPF) and broadband band pass filter (BBPF). 
  The numerical simulation results exhibit that the performance of filters is 
  improved based on the inverse design and weighting operation. For instance, 
  the bandwidth and quality factor of the optimized NBSF can reach 6.5 nm and 200.5, 
  respectively. In addition, some relatively complex transmission characteristics 
  can also be obtained by using the PPWS, for example single plasmon-induced transparency 
  (PIT)-like effect, double PIT-like effects and wavelength-tunable PIT-like effects. 
  In conclusion, GA is considered as the most efficient inverse design method 
  for our system due to less optimization time and stable performance.

\section{Device design and simulation results}

  As shown in Fig. \ref{fgr1:device}(a), our proposed PPWS consists of a MDM waveguide and MCMs. 
  The MDM waveguide and MCMs are placed on a SiO$_2$ substrate. 
  Here, the MCMs whose footprint is ${400}\times{400}$ nm$^2$ and composition is ${M}\times{N}$ square pixels
  are placed on one side of the MDM waveguide. Each pixel in the MCMs 
  can be selectively filled by Ag or air, corresponding to the logical state 
  "0" or "1", respectively. Obviously, MCMs provide a relatively broad programming 
  space which is expected to realize different functions based on the inverse design 
  technology. In the PPWS, the length and width of the MDM waveguide which transmits
  the transverse magnetic (TM) polarized light are 620 nm and 100 nm, respectively. 
  The TM mode in the MDM waveguide can be coupled into the MCMs partially and then 
  be coupled back into the MDM waveguide. The excited modes in the MCMs 
  will have impact on the optical transmission characteristics in the transmission 
  spectrum. In our simulations, the transmission spectrums are calculated by 
  using the 2-dimensional (2D) finite-difference time-domain (FDTD) method 
  (adopting Lumerical FDTD Solutions). The maximum mesh step is set as 1/4 
  of the length of single pixel in MCMs. The relative dielectric constant $\varepsilon(\omega)$ 
  of the metal (Ag) in the FDTD simulation is modeled by the Drude model with 
  $(\varepsilon_{\infty}, \omega_{p}, \gamma) = (3.7, 1.37\times10^{16} Hz, 2.7\times10^{13} Hz)$
  by the following formula \cite{johnson1972optical}:

  \begin{figure}
    \includegraphics{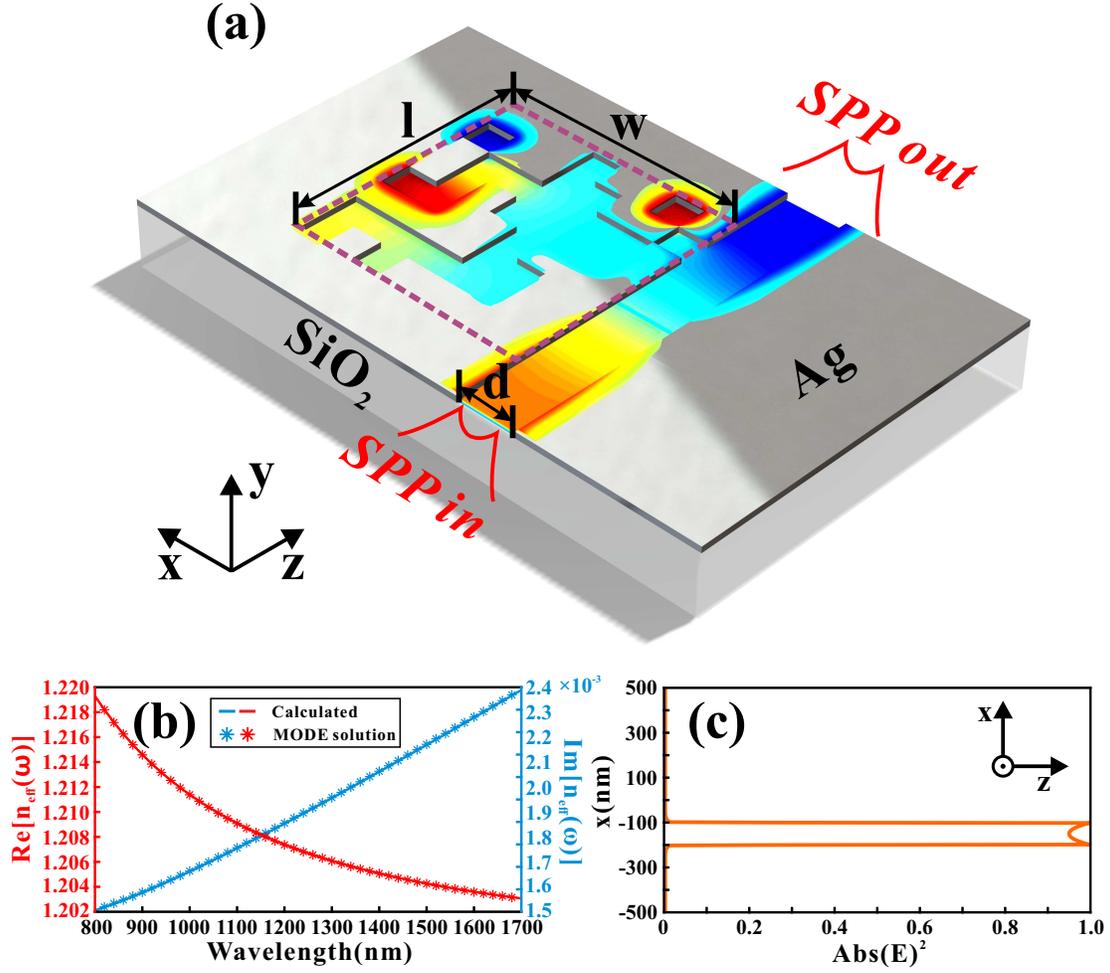}
    \caption{Structure schematic of the proposed PPWS, the effective refractive index and 
    electric field intensity distribution in MDM. (a) The schematic of MDM waveguide and MCMs. 
    (b) The real part (red) and imaginary part (blue) of $n_{eff}$. 
    The solid lines represent the theoretical calculation results and star markers 
    represent simulation results by MODE solution. 
    (c) The electric field intensity distribution of TM mode in MDM.}
    \label{fgr1:device}
  \end{figure}  

  \begin{equation}
    \varepsilon(\omega)=\varepsilon_{\infty}-\frac{\omega_{p}^2}{\omega^2+i\omega\gamma}
    \label{eqn:epsilon}
  \end{equation}
  where $\varepsilon_{\infty}$ is the interband-transition contribution to the permittivity, 
  $\omega_{p}$ is the bulk plasma frequency, and $\gamma$ is the electron collision frequency. 
  The dispersion relation of the MDM in our system is governed by the following dispersion equation:

  \begin{equation}
    \tanh(\frac{d}{2}k_{1})=-\frac{\varepsilon_1}{\varepsilon_2}\frac{k_2}{k_1}
    \label{eqn:dispersion}
  \end{equation}
  where $\varepsilon_1$ and $\varepsilon_2$ are the relative dielectric constants of the dielectric and 
  the metal, respectively. $d$ is the width of the inner dielectric. 
  $k_1$ and $k_2$ are the transverse propagation constants in the dielectric 
  and the metal, respectively, which are related to the effective refractive index $n_{eff}$ as:

  \begin{equation}
    k_i=\sqrt{n_{eff}^2k_0^2-\varepsilon_i^2k_0^2},(i=1,2)
    \label{eqn:neff}
  \end{equation}
  
  The real part and imaginary part of $n_{eff}$ theoretically calculated and numerically simulated 
  by Lumerical MODE Solutions (which has the same solution in FDTD Solutions) are plotted 
  in Fig. \ref{fgr1:device}(b). The results show that our 2D FDTD simulations is in consistent with theory. 
  Moreover, Fig. \ref{fgr1:device}(c) exhibits the electric field intensity distribution of $x$ direction 
  in MDM which is a kind of symmetric TM mode. All the transmission spectrums in our 
  simulations are calculated from 800 nm to 1700 nm in the near infrared region. 
  It should be noted that the MDM waveguide system is usually simulated by using 
  2D FDTD simulation because of the less simulation time and high accuracy 
  \cite{cao2013formation,zhan2016tunable,pannipitiya2010improved}. Although the 3D FDTD simulations 
  are more close to the practical devices, their time consumption is relatively high.
  
\begin{figure}
    \includegraphics{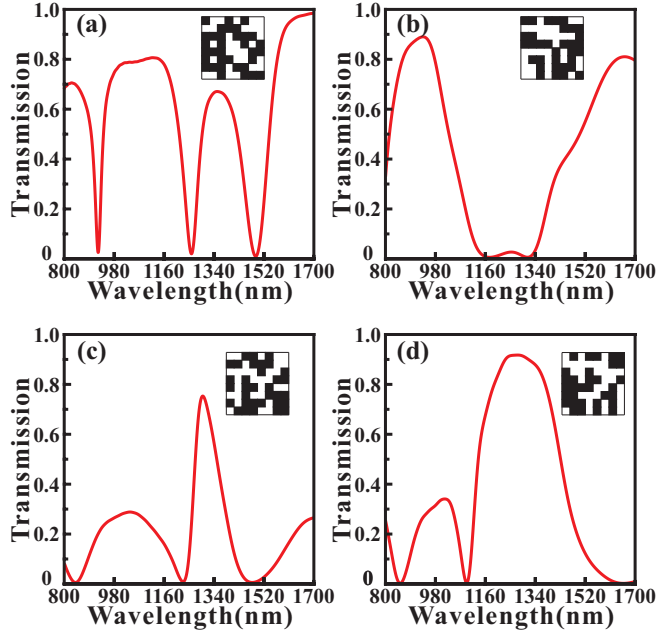}
    \caption{The simulated transmission spectrums for the MCMs with different distributions. 
    (a) The transmission spectrum that is expected to be used as a NBSF. 
    (b) The transmission spectrum that is expected to be used as a BBSF. 
    (c) The transmission spectrum that is expected to be used as a NBPF. 
    (d) The transmission spectrum that is expected to be used as a BBPF.}
    \label{fgr2:multifun}
  \end{figure}

  As mentioned above, the PPWS provides a relatively broad programming space 
  to support different functions. In practice, the MCMs can be dynamically 
  controlled by using a field-programmable gate array \cite{cui2014coding}. Here, we randomly 
  select several MCMs to exhibit the transmission characteristics of the PPWS, 
  and the simulation results are shown in Fig. \ref{fgr2:multifun}. In this simulation, 
  the MCMs whose composition is 8$\times$8 square pixels reveal that the PPWS has 
  a programming space with 2$^{64}$ distributions although some distributions 
  are senseless and repetitive for one same objective. From Fig. \ref{fgr2:multifun}(a-d), 
  it can be observed that those transmission spectrums are expected to 
  achieve multi-functional filtering, which is a fundamental and important unit 
  in the integrated plasmonic circuits. And these transmission spectrums 
  are simulated for the MCMs randomly selected from all possible distributions. 
  Obviously, it indicates that the distribution of MCMs could greatly 
  influence the output transmission spectrums and the performance metrics 
  of these filters can be further improved based on the inverse design or optimization algorithms.

\section{Multifunctional photonic device}

  \subsection{Multifunctional filter}

    In section II, we have preliminarily proved that the PPWS has potential 
    as multifunctional filters. It is expected to further improve the performance 
    metrics of the filtering based on the inverse design technology. 
    For example, we want to design a NBSF whose central wavelength 
    and 3dB bandwidth are 1300 nm and 20 nm, respectively. 
    From the beginning, we randomly generate a number of MCMs whose 
    transmission spectrums are simulated by using the 2D FDTD simulation. 
    Then, the MCMs are optimized by using the GA, PSO, SA and 
    DBS to achieve better performance, such as a narrower bandwidth. 
    GA is a highly parallel and adaptive optimization algorithm, 
    which encodes the solutions as genes to achieve evolutionary optimization. 
    Because of the complexity of gene coding, the binary coding, 
    which is suitable for our MCMs, is often used as a simplified substitution \cite{skaar1998a}. 
    The algorithmic details of the GA used here are outlined as follows: 
    (i) an initial population consisted of $N=50$ individuals are randomly generated. 
    Here, each individual in the initial population represents a distribution of 
    MCMs whose pixels are randomly initialized as "0" or "1" state. 
    It should be noted that $N$ is an important parameter of the population 
    which has influence on the convergence speed and optimization results. 
    (ii) For all $N$ MCMs, the transmission spectrums are calculated by 
    using the 2D FDTD simulation. The Euclidean distance between the 
    simulated transmission spectrum and the targeted transmission 
    spectrum is defined as the optimization objective function (loss) 
    for the GA. And it can determine whether one individual should be 
    eliminated or not in the evolution. Furthermore, the setting of 
    optimization objective function will be discussed in detail as a 
    single part at the end of this section. (iii) A new population are 
    generated by using the selection, crossover and mutation procedures. 
    In the selection process, two parent individuals are selected from 
    previous generation based on the roulette-wheel selection with 
    a gap selection ratio of 0.9 \cite{zhang2019efficient}.
    In the crossover process, the selected MCMs cross over to generate 
    a new MCMs based on the uniform crossover method with a crossover 
    probability of 0.3. In the mutation process, each pixel in the MCMs has 
    3\% probability to flip from 0 (1) to 1 (0). (iv) The newly generated 
    MCMs are evaluated to determine the GA whether to stop or not. 
    If the genetic generation reaches to 100 or the value of optimization 
    objective function remains unchanged for more than 30 times, 
    then GA stops, otherwise, proceeds to Step (ii).

    In comparison to GA, PSO is an evolution algorithm based on 
    group cooperation, which is inspired from the foraging behavior 
    of birds \cite{kennedy2002particle}. The solution (we call it "particle") of the 
    optimization problem is a "bird" in the search space. 
    All particles have a fitness value determined by 
    the optimization objective function which is the same with GA, 
    and each particle has a velocity that determines the direction 
    and distance for their flight. It should be noticed that the 
    standard PSO is suitable for the decimal problem rather than 
    binary problem so that the discrete binary PSO (DBPSO) rather 
    than the standard PSO is used to optimize the MCMs \cite{kennedy1997a}. 
    In the DBPSO, the calculation method for the velocity is 
    as same as the standard PSO. But the velocity of the 
    standard PSO directly affects the position of particle, 
    while that of DBPSO is converted to a flip probability 
    based on the sigmoid function \cite{chuang2008improved}. 
    Then, this flip probability determines whether the pixels in the MCMs are changed or not. 
    It should be noted that the velocity range, inertia weight and 
    acceleration constants of the DBPSO in the optimization are 
    set as -1$\sim$1, 1, and $c_1=c_2=1.49$ \cite{eberhart2000comparing}, respectively. SA is a typical 
    search algorithm which imitates the physical annealing in 
    the quenching based on Monte Carlo iterative solution strategy \cite{Laarhoven1987}. 
    At the initial stage, SA searches for the solution in a 
    broad optimization space by accepting a worse solution with 
    a certain probability. It can effectively alleviate 
    the local optima problem in the GA and PSO to some extent \cite{jamili2011a,yu2000a}. 
    As the temperature decreases, the solution changes in a 
    small range to speed up the convergence of SA. 
    Here, in each iteration, the MCMs are randomly generated 
    and simulated by the 2D FDTD simulation. 
    The objective function is evaluated to determine the newly 
    generated MCMs whether to be accepted or not with an 
    acceptance probability determined by the Metropolis criterion \cite{Kirkpatrick671}. 
    Here, the maximum and minimum of the temperature for the SA are 
    set as 2000 and 1$^{-18}$, respectively. And the iteration times for 
    the SA depend on the decrease rate (0.98) of temperature. 
    In addition, DBS is a nonlinear search algorithm which is 
    suitable for discrete binary image \cite{seldowitz1987synthesis}. From the beginning, 
    each pixel is traversed in turn, and the state with the 
    smaller value of optimization objective function is 
    selected and then fixed each time \cite{Shen2015}. It should be noticed that 
    the optimal solution will only approach to the direction of 
    convergence in this selection process, so there will be no 
    oscillation and rebound for the value of optimization objective 
    function. Moreover, in order to enhance the performance of PPWS, 
    we increase the time for traversing all pixels. In this article, 
    if the traversal times of the DBS reaches to maximum or the 
    value of loss remains unchanged for more than 10 times, 
    then stopping the search.
    
\begin{figure}
      \includegraphics{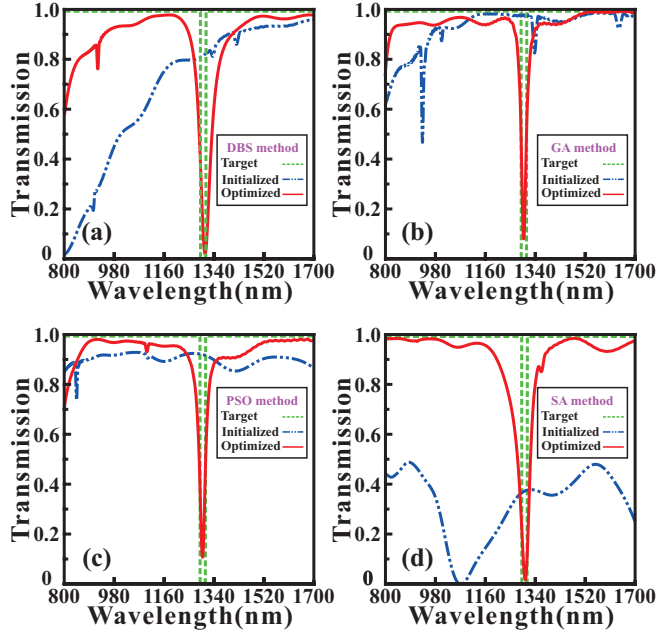}
      \caption{The initial transmission spectrums (blue dash lines) and 
      optimized transmission spectrum (red solid lines) for 
      (a) DBS, (b) GA, (c) DBPSO and  (d) SA. 
      The green dotted lines in (a), (b), (c) and (d) 
      are the targeted transmission spectrums.}
      \label{fgr3:difmethod}
    \end{figure}

    We use GA, DBPSO, SA and DBS to optimize for the MCMs in the PPWS 
    to achieve a NBSF for demonstrating the availability and 
    effectiveness of optimization algorithms. Here, the MCMs in the PPWS 
    are composed of 8$\times$8 pixels. The targeted central wavelength of the NBSF 
    is set as 1300 nm and the targeted full width at half maximum (FWHM) 
    of the transmission dip is set as 20 nm. Here, the line shape of the 
    targeted transmission spectrum is set as rectangle function. 
    The targeted transmission spectrums (green dash lines) and 
    the optimized transmission spectrums (red solid lines) are 
    shown in Fig. \ref{fgr3:difmethod}. It can be found that the transmission spectrums 
    have a notable improvement after optimizing with four optimization 
    methods since the optimized transmission spectrums are very close 
    to the targeted transmission spectrums comparing with the initial 
    transmission spectrums (blue dash line). Here, the initial 
    transmission spectrums of the DBS and SA are calculated for 
    the initial random distributions of the MCMs, while those of 
    GA and DBPSO are the best individuals in the first generation. 
    After optimizing, we can find that the minimum transmittances 
    of the transmission dips are reduced to 0.0208, 0.0812, 0.1045 and 
    0.0184 for DBS, GA, DBPSO and SA, respectively. And 
    the central wavelengths (FWHMs) of the optimized NBSFs 
    are 1307.0 nm (7.5 nm), 1299.5 nm (7.5 nm), 1299.0 nm (10.5 nm) and 
    1303.0 nm (6.5 nm) for DBS, GA, DBPSO and SA, respectively. 
    Correspondingly, the Q-factors of the optimized NBSFs can 
    reach 174.27, 173.67, 123.71 and 200.46, respectively. 
    Obviously, all optimization algorithms are convergent and 
    effective because the optimized transmission spectrums are 
    close to the targeted transmission spectrums. Even so, there 
    exist slight performance differences of the fluctuation degree 
    in the sideband of the optimized transmission spectrums. 
    The max fluctuations of the sideband in the transmission spectrums 
    optimized by the DBS, GA, DBPSO and SA are 2.28 dB, 1.32 dB, 1.48 dB 
    and 0.64 dB, respectively. Though SA has the minimum transmittances 
    and FWHM, which indicates it performs well in this simulation, 
    it is unstable in the optimization process due to the sensitivity 
    of initial condition. GA, by contrast, can get gratifying performance 
    even if it only executes once or twice, which is more stable than others.

\begin{figure}
      \includegraphics{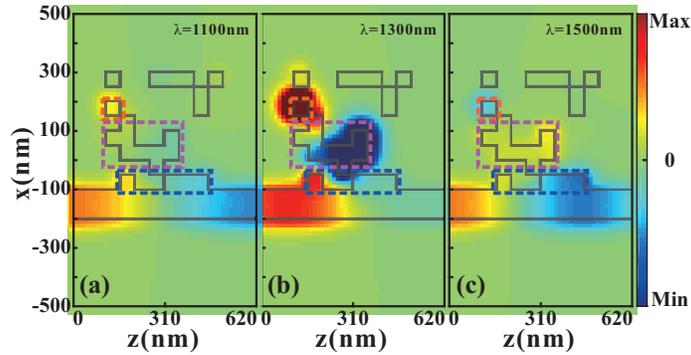}
      \caption{The distribution of the MCMs and $Re(Hz)$ for 
      the optimized NBSF at (a) 1100 nm, (b) 1300 nm, and (c) 1500 nm.}
      \label{fgr4:filterHz}
    \end{figure}

    After the optimization, the distribution of the MCMs and 
    the $Re(Hz)$ (real part) of the transmission characteristics for NBSF 
    are shown in Fig. \ref{fgr4:filterHz}. The distribution of MCMs comes from the optimization 
    results of DBS method in the Fig. \ref{fgr3:difmethod}. It can be found that for 
    the transmission points at wavelengths 1100 nm and 1500 nm, 
    the SPPs are slightly coupled into MCMs through the pixels 
    surrounded by blue boxes. Obviously, it doesn't form a stable resonant 
    mode so that SPPs pass through the MDM waveguide with low loss. 
    For the transmission dip at 1300nm, the SPPs are coupled into 
    the MCMs to form a strong resonant mode between the orange zone 
    and magenta zone. Thus, an obvious transmission dip emerges 
    in the transmission spectrum.

\begin{figure}
      \includegraphics{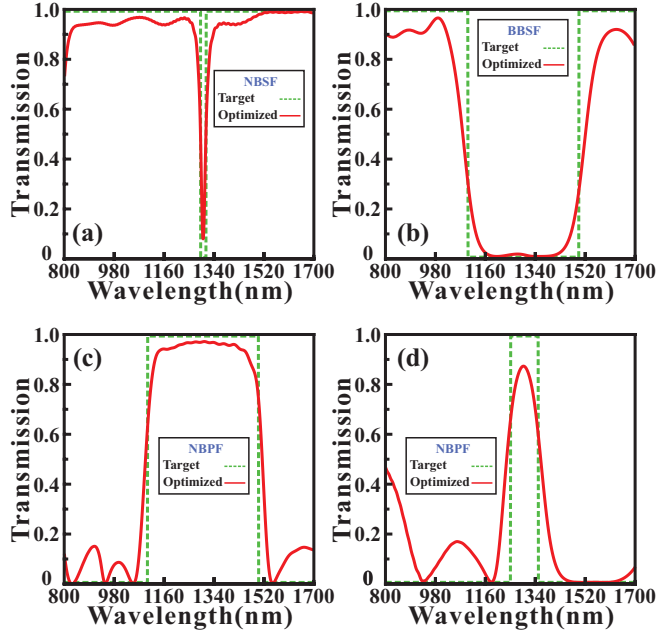}
      \caption{The transmission spectrums optimized by GA for (a) NBSF, (b) BBSF (c) BBPF 
      and (d) NBPF. The green dotted line is the targeted transmission spectrum. 
      Red solid line is the optimized transmission spectrum.}
      \label{fgr5:otherfilter}
    \end{figure}

    Then, we use GA to optimize for other three types of filters (BBSF, BBPF and NBPF), 
    and corresponding simulation results are shown in Fig. \ref{fgr5:otherfilter}. 
    For contrast, the optimized results for the NBSF are shown in Fig. \ref{fgr5:otherfilter}(a). 
    Fig. \ref{fgr5:otherfilter}(b) shows the optimized transmission spectrum for the BBSF, 
    which has a wide stopband from 1100 nm to 1500 nm. 
    It can be found that the transmittances are less than 0.01 in the stopband and 
    higher than 0.8 in the passband, indicating this BBSF can achieve 
    promising filtering effects. In comparison to the NBSF, the stopband of the BBSF 
    is not enough smooth. The reason for this phenomenon is attributed to 
    the optimized MCMs have redundant pixels, which may lead to the deterioration of 
    performance in the stopband. In addition, Fig. \ref{fgr5:otherfilter}(c) shows the optimized 
    transmission spectrum for the BBPF that has a broad passband from 1100 nm to 1500 nm. 
    Here, the maximum transmittance and FWHM of the BBPF reach 0.9718 and 423 nm, respectively. 
    And the maximum fluctuation degree in the passband of the BBPF is only 0.14 dB, 
    which indicates that the passband is very flat. Moreover, 
    we design a NBPF whose 3 dB bandwidth is 133.5 nm, 
    and it’s optimized transmission spectrum is shown in Fig. \ref{fgr5:otherfilter}(d). 
    It can be observed that the maximum transmittance of the transmission peak 
    can reach 0.8668 at 1296.5 nm.  Obviously, the bandwidth and sideband inhibition of 
    the NBPF can’t be optimized as good as the NBSF. The reason for this phenomenon is 
    related to the loss in the MCMs. In other words, the MCMs is easily used to inhibit 
    rather than promote the transmission of specific wavelengths so that the NBSF 
    have better performance than NBPF. In conclusion, the PPWS is demonstrated to 
    be able to realize multiple optical filtering functions with the remarkable results. 
    And we do not need to make any effort to analysis the resonant mode in the system ahead.

    Next, two key elements which have significant influence on the optimal performance 
    of filtering functions are discussed. Firstly, it is clear that the similarity between 
    the simulated transmission spectrum and targeted transmission spectrum determines 
    the performance after optimization. Here, we propose a simple but effective method 
    named "weighting operation" to calculate the similarity for improving 
    the performance metrics. As mentioned above, the Euclidean distances between 
    the simulated transmission spectrums and targeted transmission spectrums are regarded as 
    the optimization objective function (loss). Based on this, we use the weighting factor 
    to calculate the Euclidean distance. The details are outlined as follows: 
    (i) dividing the wavelength range of the transmission spectrums into two parts 
    according to the importance. (ii) One part is called "The central wavelength range (TCWR)", 
    which relates to the wavelength range of stopband or passband we focused on. 
    The other is called "The else wavelength range (TEWR)", 
    which has negligible influence on performance metrics. 
    (iii) In each iteration, the loss $L$ is calculated by the following formula:

    \begin{equation}
      L=w_{center}{\times}f(T_{center}^{sim}-T_{center}^{tar})+w_{else}{\times}f(T_{else}^{sim}-T_{else}^{tar})
      \label{eqn:weighting}
    \end{equation}
    where $w_{center}$ is the weighting factor of TCWR, $w_{else}$ is the weighting factor of TEWR. 
    $f$ is the function of calculating Euclidean distance. $T_{center}^{sim}$ and $T_{center}^{tar}$ represent 
    the simulated transmission spectrum and targeted transmission spectrum in TCWR. 
    While $T_{else}^{sim}$ and $T_{else}^{tar}$ are the simulated transmission spectrum and 
    targeted transmission spectrum in TEWR.

\begin{figure}
      \includegraphics{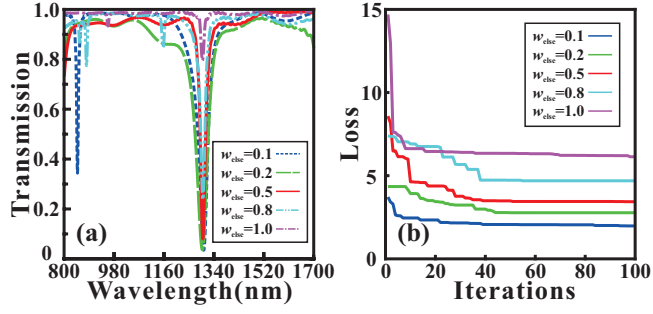}
      \caption{(a) The optimized transmission spectrums for the NBSF with 
      different weighting factors $w_{else}$. (b) The variations of loss value 
      for different weighting factors $w_{else}$. In all settings, $w_{center}$ is set as 1.}
      \label{fgr6:weight}
    \end{figure}

    In order to analyze the influence of the weighting operation, 
    we use GA to optimize for the MCMs in the PPWS to achieve a NBSF. 
    Here, the MCMs are composed of 8$\times$8 pixels. The targeted central wavelength of 
    the NBSF is set as 1300 nm and the FWHM of the transmission dip is set as 20 nm. 
    TCWR includes the wavelength range from 1290 nm to 1310nm, 
    and the rest of wavelength range belongs to TEWR. 
    The optimized transmission spectrums for different weighting factors $w_{else}$ of 
    TEWR are shown in Fig. \ref{fgr6:weight}(a). After the optimization, 
    we find that the loss values with different weighting factors $w_{else}$ 
    have a significant decline in Fig. \ref{fgr6:weight}(b). The loss values decreases from 
    3.72 to 1.98 with a 46.77\% decline ($w_{else}=0.1$), 4.35 to 2.77 with 
    a 36.32\% decline ($w_{else}=0.2$), 8.58 to 3.44 with a 59.91\% decline ($w_{else}=0.5$), 
    7.39 to 4.69 with a 36.54\% decline ($w_{else}=0.8$) and 14.69 to 6.14 with 
    a 58.20\% decline ($w_{else}=1.0$), respectively. And the loss value has 
    continuously decline with weighting factor $w_{else}$ reducing, indicating that 
    the weighting factor has a significant influence on the optimization effects. 
    In Fig. \ref{fgr6:weight}(a), it can be found that the minimum transmittance at transmission dip 
    is only 0.7968 when $w_{else}=1.0$, which is still large for filter even if 
    the loss value has a steep and quick decline in Fig. \ref{fgr6:weight}(b). 
    As weighting factor $w_{else}$ decreases, the minimum transmittance at 
    transmission dip reduces to 0.2389, 0.0812, 0.0376 and 0.0360, respectively. 
    However, it can be observed that the sideband has more violent fluctuations 
    when $w_{else}$ decreases from 1.0 to 0.1. Besides, the 3dB bandwidths are 
    14.5 nm, 7.5 nm, 10.5 nm and 8 nm for $w_{else}=0.8$, $w_{else}=0.5$, $w_{else}=0.2$ and 
    $w_{else}=0.1$, respectively. As $w_{else}$ decreases, the fluctuations in TEWR 
    become more and more strong because the contribution of TEWR becomes smaller. 
    At the same time, the significance of TCWR is amplified equivalently 
    so a small transmittance at transmission dip can be achieved when $w_{else}$ is decreased. 
    In this article, we use a weighting factor of $w_{else}=0.5$, 
    which guarantees a balance between bandwidth, minimum transmittance and 
    fluctuations of sideband. Obviously, it is better to dynamically adjust 
    the weighting factor according to the specific application and optimization results of the PPWS.

    The other key element for the optimization of the PPWS is the density of pixels in MCMs. 
    Obviously, the diversity of MCMs determines the practical functions and performance 
    of the PPWS. Mathematically, the permutation and combination of MCMs grows exponentially 
    as the density of pixels in MCMs increasing. It should be noted that the optimization algorithms 
    we used are suitable for global optimization task but their ability are not infinite. 
    Therefore, it is necessary to choose an appropriate density of pixels in MCMs, 
    which can not only ensure the diversity of MCMs but also maintain the efficiency of 
    the optimization algorithms. In order to analyze the influence of density on 
    the optimization performance, we use all algorithms to optimize for a NBSF with four different densities.

\begin{figure}
      \includegraphics{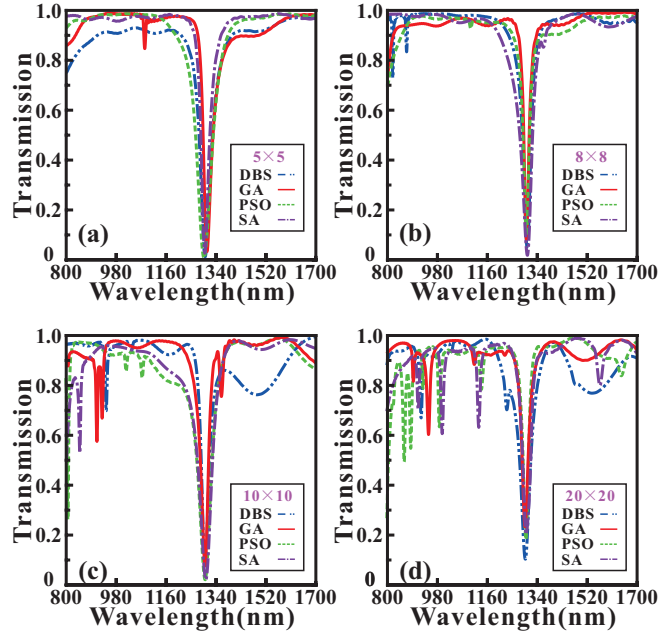}
      \caption{The optimized transmission spectrums of the NBSF with the MCMs 
      whose densities are (a) $5\times5$ pixels (b) $8\times8$ pixels (c) $10\times10$ pixels 
      (d) $20\times20$ pixels. Blue lines, red lines, green lines and purple lines 
      are the transmission spectrums optimized by MDBS, GA, DBPSO, and SA, respectively.}
      \label{fgr7:density}
    \end{figure}

    The targeted central wavelength of the NBSF is set as 1300 nm and 
    the FWHM of the transmission dip is set as 20 nm. The optimized transmission spectrums 
    of the NBSF with the MCMs whose densities are $5\times5$ pixels, $8\times8$ pixels, 
    $10\times10$ pixels and $20\times20$ pixels are shown in Fig. \ref{fgr7:density}(a)-(d), respectively. 
    The important performance metrics of the optimized transmission spectrums 
    for different densities are exhibited in Table \ref{tbl:difMCMs}.  Here, $T_{min}$ and $L_c$ are 
    the minimum transmittance and the central wavelength of the transmission dip. 
    It can be found that $T_{min}$ and 3dB bandwidth have remarkable declines as 
    the density of MCMs increases. And the minimum transmittance and 3dB bandwidth 
    for $20\times20$ pixels perform the worst in comparison to other densities. 
    It can not be neglected that the fluctuations in the sideband for $20\times20$ pixels 
    are extremely violent, which may restrict the applications of filtering function. 
    The reason for this phenomenon is related to that the solution space of $20\times20$ pixels 
    is too broad for optimization algorithm to search for the global optimal solution. 
    In addition, it should be noted that the FDTD simulation for high-density MCMs will take much more time to 
    converge as the computational complexity of the high-density MCMs is increased significantly. 
    The choice of density for the MCMs depends on the practical requirements. Moreover, 
    it can be found that the performance differences between $5\times5$ pixels, $8\times8$ pixels and 
    $10\times10$ pixels are not obvious. In comparison to $5\times5$ pixels and $10\times10$ pixels, 
    the inhibition of the sideband for $8\times8$ pixels performs better because it is smoother. 
    The 3dB bandwidth and $T_{min}$ for $5\times5$ pixels are more competitive, 
    but the offset of central wavelength is relatively large, 
    which limits the precise filtering. Considering the diversity and computational efficiency, 
    we select the MCMs with $8\times8$ pixels to achieve most of the functionality in this article.

    \begin{table}
      \caption{Performance in different density of MCMs}
      \label{tbl:difMCMs}
      \begin{tabular}{clccc}
        \hline
        Density & Method  & $L_c$ & $T_{min}$ & 3dB bandwidth \\
        \hline
        \multirow{4}{*}{$5\times5$}
         & DBS     & 1300.5 & 0.0124 & 6.0 \\
         & SA      & 1299.0 & 0.0267 & 6.5 \\
         & DBPSO   & 1296.5 & 0.0071 & 6.0 \\
         & GA      & 1307.0 & 0.0330 & 7.0 \\
        \hline
        \multirow{4}{*}{$8\times8$}
        & DBS     & 1302.5 & 0.0441 & 7.5 \\
        & SA      & 1303.0 & 0.0184 & 6.5 \\
        & DBPSO   & 1299.0 & 0.1045 & 10.5 \\
        & GA      & 1299.5 & 0.0812 & 7.5 \\
        \hline
        \multirow{4}{*}{$10\times10$}
        & DBS     & 1299.5 & 0.0657 & 9.5 \\
        & SA      & 1302.5 & 0.0185 & 8.0 \\
        & DBPSO   & 1302.0 & 0.0206 & 8.5 \\
        & GA      & 1299.5 & 0.0737 & 10.0 \\
        \hline
        \multirow{4}{*}{$20\times20$}
        & DBS     & 1299.5 & 0.1067 & 16.5 \\
        & SA      & 1302.5 & 0.2064 & 23.5 \\
        & DBPSO   & 1301.5 & 0.1855 & 13.5 \\
        & GA      & 1298.5 & 0.2364 & 19.5 \\
        \hline
      \end{tabular}
    \end{table}

  \subsection{PIT-like effect}
\begin{figure}
      \includegraphics{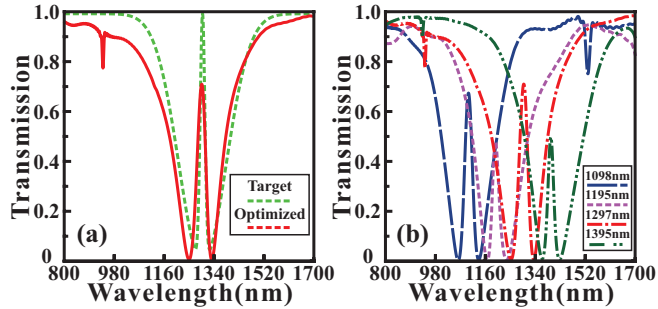}
      \caption{(a) The transmission spectrum optimized for a single PIT-like effect. 
      The green dotted line is the targeted transmission spectrum. 
      Red solid line is the optimized transmission spectrum. 
      (b) Wavelength-tunable PIT-like effects with different 
      central wavelength 1100 nm, 1200 nm, 1300 nm and 1400 nm.}
      \label{fgr8:singlePIT}
    \end{figure}

    In the previous section, we show that the PPWS can achieve a variety of 
    filtering functions, but these functions are relatively simple. Actually, 
    the PPWS can be programmed to realize advanced functions because of 
    the diversity of MCMs. In addition, this programmable waveguide system can 
    also be used to achieve wavelength-tunable optical devices based on 
    optimization algorithms and the MCMs. Here, we use the GA, whose 
    comprehensive performance is the best in the previous section, 
    to optimize for the MCMs to achieve the PIT-like effects in the 
    transmission spectrum. The optical characteristic of the PIT-like effects 
    includes a transmission peak located between two transmission dips, 
    which has great applications in optical switches, optical data storage 
    and slow light \cite{han2011plasmon,wang2012dispersionless}. In this 
    FDTD simulation, the MCMs are composed of 8$\times$8 pixels. 
    The targeted FWHM of the transmission peak is set as 20 nm, 
    while the targeted FWHM of the whole transmission dip is set as 200 nm. 
    The transmission spectrums optimized for the single PIT-like effect are 
    shown in Fig.~\ref{fgr8:singlePIT}. Obviously, it can be observed that the single PIT-like 
    effect whose center wavelength is 1297 nm is shown in Fig. \ref{fgr8:singlePIT}(a), 
    and the targeted central wavelength of the transmission peak is set as 1300 nm. 
    The transmittance of the transmission peak reaches to 0.7115, 
    while the transmittances of the transmission dips are 0.0084 (1251.5 nm) 
    and 0.0086 (1333.5 nm), respectively. As a critical parameter to evaluate 
    the performance of the PIT-like effect, the Q-factor, is calculated by the following formula:

    \begin{equation}
      Q=\frac{f_0}{FWHM}
      \label{eqn:Q}
    \end{equation}
    where $f_0$ is the center wavelength of the single PIT-like effect. 
    After the GA optimization, the $FWHM$ in Fig. \ref{fgr8:singlePIT}(a) is 35.5 nm so that 
    the Q-factor reaches to 36.54. In fact, the Q-factor is lower in the 
    metallic systems due to the ohmic damping \cite{hu2018comparison}. What's more, 
    there is no need to analyze the complex coupling mechanism in a 
    metallic resonator to achieve the single PIT-like effect. 
    In our simulation, we just focus on using efficient optimization algorithms 
    and reasonable physical conditions to obtain the targeted optical characteristics.

    In addition to the PIT-like effects at specific wavelength, 
    the wavelength-tunable PIT-like effects are expected to be applied 
    in highly integrated optical circuits due to flexibility \cite{han2015dynamically,zhang2017tunable}. 
    For example, it has been demonstrated that the tunable PIT-like effects 
    can be applied in optical modulators, switches, sensors, slow light 
    and so on \cite{wang2015tunable}. Fig. \ref{fgr8:singlePIT}(b) shows the wavelength-tunable PIT-like 
    effects based on the PPWS with different central wavelengths. 
    In the optimization process, we only need to adjust the central position 
    of the targeted PIT-like effects and keep other parameters fixed. 
    The center (targeted) wavelengths of the tunable PIT-like effects 
    are 1098 nm (1100 nm), 1195 nm (1200 nm) and 1395 nm (1400 nm), respectively. 
    The maximum deviation between the center wavelength and targeted wavelength is only 5~nm. 
    The corresponding FWHMs (Q-factors) of the tunable PIT- like effects are 
    31.5 nm (34.86), 27 nm (44.26) and 24 nm (58.13), respectively. And the extinction ratios 
    between the transmission peak and two transmission dips are 22.53 dB (left) and 20.38 dB (right) for 
    the PIT-like effect at 1098 nm, 20.52 dB (left) and 18.93 dB (right) for 
    the PIT-like effects at 1195 nm, 16.55 dB (left) and 18.22 dB (right) for 
    the PIT-like effects at 1395 nm, respectively. Obviously, the FDTD simulation results show 
    that the tunable PIT-like effects can be obtained in the wavelength range from 1100 nm to 1400 nm, 
    which demonstrates that the PPWS has great flexibility and controllability based on the inverse design.

    \begin{figure}
      \includegraphics{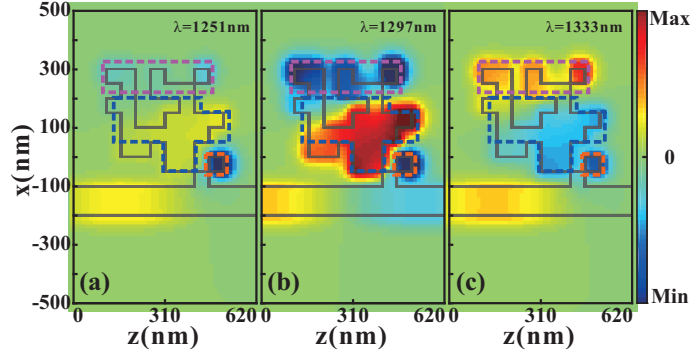}
      \caption{The distribution of the MCMs and PIT-like effects whose central wavelength 
      is at 1297nm. The field distribution of $Re(Hz)$ 
      at (a) 1251.5 nm, (b) 1297 nm, (c) 1333.5 nm.}
      \label{fgr9:PITHz}
    \end{figure}

    For the single PIT-like effect, the distribution of the MCMs and 
    corresponding electromagnetic field distributions are shown in Fig. \ref{fgr9:PITHz}. 
    Based on the magnetic field distributions, we provide a qualitative explanation 
    for the single PIT-like effect. It can be found that SPPs are confined in these regions 
    which are marked by the yellow rectangular box and red rectangular box, 
    leading to the transmission dips at 1251.5 nm and 1333.5 nm, respectively. 
    For the transmission peak at 1297 nm, SPPs are coupled into yellow zone through 
    the red zone to form a resonant mode. And the phase is same when SPPs are coupled back into 
    the strip waveguide so there is a transmission peak. It should be noted that it is difficult 
    to use the bright-dark mode coupling mechanism or doublet of dressed states to explain 
    the PIT-like effects due to the irregular shapes of MCMs \cite{wang2014analogue}.

    \begin{figure}
      \includegraphics{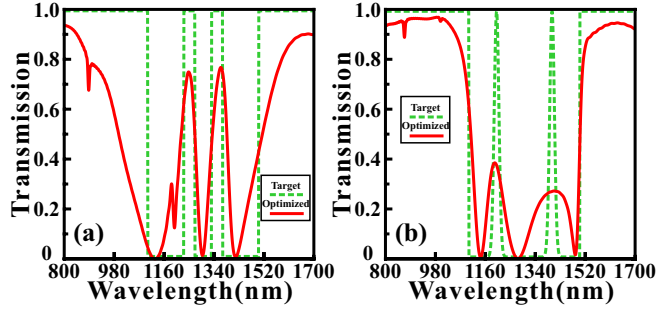}
      \caption{(a) The simulated transmission spectrums of the double PIT-like effects 
      with higher transmittances. The targeted central wavelengths are set as 1250 nm and 1350nm. 
      The targeted FWHMs of transmission peaks are both 40 nm. 
      (b) The simulated transmission spectrums of the double PIT-like effects with 
      lower transmittances. The targeted central wavelengths are set as 1200 nm and 1400 nm. 
      The targeted FWHMs of transmission peaks are both 20 nm. 
      The green dotted lines are the targeted transmission spectrums. The red solid lines are 
      the simulated transmission spectrums.}
      \label{fgr10:doublePIT}
    \end{figure}

    Finally, we increase the complexity of the targeted transmission spectrum to achieve 
    the double PIT-like effects. The simulated transmission spectrums are shown in Fig. \ref{fgr10:doublePIT}. 
    Here, Fig. \ref{fgr10:doublePIT}(a) shows an ideal double PIT-like effects with higher transmittances, 
    whose peaks reach to 0.7502 (left channel) and 0.7684 (right channel). By contrast, 
    Fig. \ref{fgr10:doublePIT}(b) shows the double PIT-like effects with lower transmittances in 
    the transmission peaks. However, the optimization and matching effects of sideband 
    in Fig. \ref{fgr10:doublePIT}(b) are better than those in Fig.~\ref{fgr10:doublePIT}(a). It should be noted that 
    the sampling points in wavelength range are equidistant in our simulation. 
    But the critical wavelength range that determines the optimization performance 
    is relatively small. Although we use the weighting operation to eliminate 
    the influence from sideband, there still are problems to be solved to make 
    the simulation results match the targeted spectrums perfectly. Even so, 
    it can be observed that the PPWS can be potentially applied to achieve 
    more complex functions from Fig. \ref{fgr9:PITHz} and Fig. \ref{fgr10:doublePIT}.

\section{Conclusion}

In conclusion, we propose a PPWS based on MCMs and the inverse design. 
Several optimization algorithms, such as GA, SA, DBS and DBPSO are applied in 
the design of the MCMs to achieve different functions. The weighting operation can 
improve the optimization performance. And the FDTD simulation results demonstrate 
that the optimized PPWS can not only obtain simple functions (such as filters), 
but also achieve more complex transmission characteristics (for example 
single PIT-like effect, double PIT-like effects and wavelength-tunable PIT-like effects). 
Although the perfect matching between the targeted transmission spectrum and
optimized transmission spectrum is still difficult to achieve. 
These results indicate that the PPWS combined with the inverse design has 
great potential to realize multifunctional plasmonic devices with low cost 
and high efficiency. There is no need to spend much time on considering the 
shapes of resonators because the MCMs in the PPWS are reprogrammable. 
In comparison to the previous works, our proposed PPWS not only provides 
a general framework for obtaining an effective, flexible and compact 
plasmonic device but also shows the applications of the inverse design on photonic devices.

\begin{acknowledgement}

This work was supported by the National Natural Science Foundation of 
China (61705015, 61625104, 61431003, 61821001); Fundamental Research 
Funds for the Central Universities (2019RC15, 2018XKJC02); 
National Key Research and Development Program of China 
(2019YFB1803504, 2018YFB2201803, 2016YFA0301300); 
Fund of State Key Laboratory of IPOC (BUPT) (NO. IPOC2020ZT08), 
P. R. China; Beijing Municipal Science and Technology Commission (Z181100008918011).

\end{acknowledgement}





\bibliography{acs_ref}

\end{document}